\documentclass[11pt,twoside]{article}   
\usepackage{epsfig}
\title{A Gedanken Experiment For Gravitomagnetism}

\author {Stanley L. Robertson\footnote{stan.robertson@swosu.edu, Physics Dept.,
Southwestern Okla. State Univ., Weatherford, OK 73096}}

\begin{document}

\maketitle                 

\begin{abstract}
A gedanken experiment implies the existence of gravitomagnetism
and raises a question about what we know about the weak-field
limit of the gravitomagnetic field of General Relativity.
\end{abstract}

\section{The Experiment}
Imagine positive charge +Q essentially fixed to the origin of
coordinates of system $S'$ by virtue of having a large rest mass,
$M_o$. Let a positive charge, q, of mass $m_o~(<< M_o)$ be placed
distance $y'$ from Q in $S'$ and experience the influence of a
Coulomb force, ${\bf F'} = kQq{\bf j}/y'^2$. The amounts of charge
are chosen such that their repulsion just balances their
gravitational attraction ${\bf F'} = -GM_0m_0{\bf j'}/y'^2$.
Having set up this nice balance, you depart along the $-x'$-axis
at speed $-v$.

\begin{figure}
\epsfig{figure=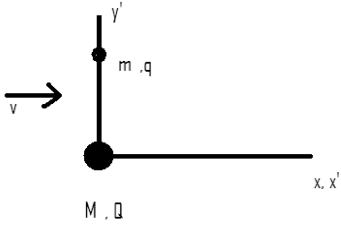,angle=0,width = 5cm, height=6.25cm}
\caption {Here $M>>m$ is located at the origin of coordinates of
the $S'$ system, which moves at speed $v$ down the common x,x'
axes relative to system $S$. Masses $M$ and $m$ are positively
electrically charged with charges Q and q, respectively. Electric
repulsion and gravitational attraction are balanced, leaving both
objects at rest with no net force between them in $S'$. In system
$S$, electric, magnetic, gravitational and gravitomagnetic forces
act.}
\end{figure}%

Obviously, the masses will remain balanced from your new
perspective in system $S$. But in your now moving frame it appears
that both electric and magnetic forces act on $q$ such that a
magnetic field ${\bf B} = {\bf v \times E}/c^2$ is created by the
motion of Q relative to you, where ${\bf E} =\gamma kQ{\bf
j}/y'^2$ is the (foreshortened) electric field at q, due to Q.
Here $\gamma =1/\sqrt{1-v^2/c^2}$. The sum of electric and
magnetic forces as calculated in $S$ is ${\bf F} = q{(\bf E + u
\times B})$, where ${\bf u}$ is the velocity of q relative to S
and, in this case ${\bf u} = {\bf v}$. This leads to combined
electric and magnetic forces of
\begin{equation}
{\bf F} = \frac {\gamma kQq{\bf j}}{y^2}(1-v^2/c^2)
\end{equation}
Clearly this is a weaker force than that observed in the original
system $S'$ and if gravity were not present it would lead to a
slower separation of $Q$ and $q$ as they repelled each other. From
$S$, you would see the separation occur in a time dilated way. But
now consider the gravitational attraction between masses $M$ and
$m$. The fact that you chose to go away and leave this balanced
arrangement of electric and gravitational forces can surely have
no bearing on its subsequent behavior. To remain motionless as
observed in system $S$, the total force of gravitational origin
between the objects in system S must surely be
\begin{equation}
{\bf F} = \frac {-\gamma GM_0m_0{\bf j}}{y^2}(1-v^2/c^2)
\end{equation}
And as originally balanced in $S'$, $GM_om_o = kQq$. The part of
the force in Eq. (2) that is velocity dependent can be considered
to be a gravitomagnetic force. In the S frame it must exist and
must, and does, balance the ordinary magnetic force. One would
expect that Eq. (2) could be obtained by analogy with the
electromagnetic case. $M$ would be the source of a $g$ field as
observed in $S$ to be ${\bf g} = -\gamma GM_0{\bf j}/y^2$ and also
a source of a gravitomagnetic field ${\bf B_g}= {\bf v \times
g}/c^2$ in which $m$ moves. Then if the sum of gravitostatic and
gravitomagnetic forces would be
\begin{equation} {\bf F} = m_0{(\bf g + u \times
B_g})
\end{equation}
we would obtain Eq. (2) without further ado.

But the weak-field limit of General Relativity produces a
gravitomagnetic field that is
\begin{equation}
{\bf B_g}= 4{\bf v \times g}/c^2
\end{equation}
(see, e.g. ref [1]) which would lead to gravitomagnetic forces
four times stronger than required to balance the ordinary magnetic
effect. The force of gravitational origin in Eq. (3) would become
\begin{equation}
{\bf F} = \frac {-\gamma GM_0m_0{\bf j}}{y^2}(1-4v^2/c^2)
\end{equation}
I leave the reader to ponder where this goes wrong.

\section{The Free-Fall Problem in General Relativity}
To see what gravitomagnetic effects we should expect, let us
dispense with the electric charges on M and m ($Q=q=0$) and just
consider the free fall of $m$ toward $M$ down the $y'$-axis
starting from rest. In either frame, S or $S'$, the small test
mass, $m$, would follow a geodesic. In the weak field limit of
$S'$ and in isotropic coordinates the metric is
\begin{equation}
ds^2=(1-2\phi)c^2 dt'^2 -(1+2\phi)(dx'^2+dy'^2+dz'^2)
\end{equation}
where $\phi = GM_o/c^2r'$. A Lorentz transform to frame $S$
[$x=\gamma(x'+vt')~~~t=\gamma(t+vx'/c^2)~~~ y=y'~~~z=z'$] yields
\begin{equation}
ds^2=(1-2\phi\gamma^2(1+v^2/c^2))c^2 dt^2
-(1+2\phi\gamma^2(1+v^2/c^2))dx^2 -(1+2\phi)(dy^2+dz^2)
+(8\gamma^2\phi v/c) dxcdt
\end{equation}
where, at the instant depicted in Figure 1, $r'=y'=y$. The last
terms with $dxdt$ arise from the Lorentz transformation. The
coefficient of $8$ is exactly 4X larger than expected from
Newtonian theory combined with special relativity. It partially
describes the ``gravitomagnetic'' effect. Of particular interest
is that $g_{01}= 4\gamma^2\phi v/c$ can be considered as arising
from a vector potential ${\bf A_g} = (4\gamma^2G/c^2)\int{ ({\bf
J}/r') dV}$ where ${\bf J} = \rho {\bf v}$ is the mass current
density and the gravitomagnetic field is taken to be ${\bf
B_g=\nabla \times A_g}$. But this term with its extra factor of 4
cannot completely describe all of the gravitomagnetic effects, for
we have seen that that the assumption that it does leads to an
unacceptable result for our gedanken experiment.

In the weak-field limit, we have $\det{g} \approx -1$, neglecting
terms involving $\phi$. We can use the inverse of the Minkowski
metric $\eta^{ij}$ to raise indexes. The geodesic equation for the
y-component of motion of $m$ is
\begin{equation}
\frac{d^2y}{c^2dt^2} \approx -(\Gamma^2_{00} +
\Gamma^2_{11}v_x^2/c^2+2\Gamma^2_{01}v_x/c)
\end{equation}
Taking $g^{22}= \eta^{22} = -1$, there follows
$\Gamma^2_{00}=\Gamma^2_{11}=-\gamma^2(1+v^2/c^2)(\partial
\phi/\partial y)$ and $\Gamma^2_{01}=(2\gamma^2v/c)(\partial
\phi/\partial y)$. Then with a little algebra and using $v_x=v
>> 0, v_z=0, v_y \approx 0$, we obtain
\begin{equation}
\frac{d^2y}{c^2dt^2} = \gamma^2\frac{\partial \phi}{\partial
y}[(1+v^2/c^2)^2-4v^2/c^2]= (1-v^2/c^2)\frac{\partial
\phi}{\partial y}
\end{equation}
which is exactly the same result that we get from Eq. (2) if the
inertial mass on which ${\bf F}$ acts is $\gamma m_0$. Eq. (9) is
a geodesic equation that is correct to order $v^2/c^2$. What
happens in Eq. (9) is that the term $\gamma^2 (1+v^2/c^2)^2$
manages to cancel $3/4$ of the effect of the factor of 4 in the
gravitomagnetic field term. It would have been very surprising to
have obtained any different result, for that would have left the
general relativistic approach in conflict with special relativity.
From the standpoint of special relativity, the freely falling mass
could serve as an interval timer, with the interval corresponding
to the time to fall some particular small distance down the
y'-axis. As observed in $S$, the interval would be dilated and
that is exactly what is described by Eqs. (2) or (9). Apparently
magnetic and gravitomagnetic forces provide mechanism for
enforcing time dilation.

The reason that the combination of Eqs. (3) and (4) fails to yield
Eq. (9) and correctly describe the free-fall problem is that they
are only correct to order $v/c$. They have been combined to
produce terms of order $v^2/c^2$, but apparently not all of them.
Under these circumstances, it would also be expected that they
would not correctly predict all gravitomagnetic effects to order
$v^2/c^2$.

\section{Special Relativity}
The gedanken experiment can be made more interesting by allowing m
to orbit M in a circular orbit of radius b as observed in $S'$. As
observed in $S$, the orbital time period will be dilated and it
can be shown from the kinematic equations of special relativity
that the orbit diameter will be the usual 2b along y-axis, but be
contracted to $2b/\gamma$ along the x-axis. Further, a special
case in agreement with special relativity can be obtained from Eq.
(8) using $v_x=v\pm u$, where $u$ would be the speed of m relative
to M, as observed in S, when m is on the y-axis. Neglecting terms
in $u^2$, the resulting acceleration of m for these positions
would be
\begin{equation}
\frac{d^2y}{c^2dt^2} \approx \frac{\partial \phi}{\partial
y}(1-v^2/c^2 \pm 2uv/c^2)
\end{equation}
The last term of the equation accounts for the small difference of
speeds of m and M relative to reference frame $S$.

\section{Laser Lunar Ranging}
The foregoing ruminations have been motivated by recent
discussions of laser lunar ranging. Murphy, Nordtvedt and Turyshev
[2] have recently published new results that claim to have
verified Eq. (4) as a source of perturbing acceleration acting on
the lunar orbit. It is claimed that they have established this
weak-field limit prediction of General Relativity to an accuracy
of about 0.1\%, which is far better confirmation than expected
from the Gravity Probe B experiment. Ciufolini [3] has pointed out
that the result of Murphy et al. is merely a coordinate dependent
effect. Their results were obtained relative to a solar system
barycenter frame (hereafter SSB) and thus could be transformed
away. Kopeiken, on the other hand, says [4] that there should be
no observable effect at all relative to the SSB.

While the gravitomagnetic field due to the sun's rotation cannot
be transformed away, it is much too small to have a measurable
effect on the lunar orbit. If a gravitomagnetic effect exists
relative to the SSB it must be associated with the motions of
earth and moon relative to the SSB. For analysis of the lunar
motions, the orbital motion of earth would be regarded as the
cause of a $\bf B_g$ field that is some four orders of magnitude
larger than that due to solar rotation at the position of the
moon.

At the time of a new moon, the situation would correspond to the
masses m and M positioned as shown in Figure 1, with the sun
located a long way up the y-axis. System $S'$ would be comoving
with the earth while system $S$ would remain at rest relative to
the SSB. Thus whatever would be observed in S should be the same
as what is measured relative to the SSB for this particular
configuration. Considering the outcome of the previous gedanken
experiment, the major effects that we should observe would be time
dilation and length contraction. For a lunar orbit radius of $3.8
\times 10^8~m$ and lunar velocity of $\sim 30~km/s$, the
contraction of the lunar orbit parallel to the nearly joint
earth-moon orbital motion about the sun would be about 1.9 m. This
would be observed in frame S of Fig. 1, but not in the SSB frame.

In view of the gedanken experiment and Eqs. (2), (9) and (10), we
should not expect the combination of Eqs. (3) and (4) to be
confirmed in the SSB frame, notwithstanding the apparent
confirmation by Murphy, Nordtvedt and Turyshev [1]. They
transformed the LLR to the SSB frame and then fitted the lunar
trajectory using multiple adjustable parameters. In recent years
the amplitude of the dominant gravitomagnetic oscillation has been
reported as $6.5~m$ [1], $9.3~m$ [5] and $5.3~m$ [6]. Although the
residuals of these curve fits to the lunar range observations are
typically less than one cm, the oscillation amplitudes are
obviously much more variable. Until the statistical errors of
determination of the oscillation amplitude exclude the
alternatives considered here, the claim that Eq. (4) has been
confirmed is unwarranted.

Finally, we should note that calculating gravitomagnetic effects
relative to a geocentric frame removes several opportunities for
error. In this frame the moon moves in a gravitomagnetic field due
to the apparent motion of the sun. Shahid-Saless has calculated
the expected amplitude of gravitomagnetic lunar orbit oscillations
to be about $3 ~cm$ in the geocentric frame.


\begin{thebibliography}{}

\bibitem{} Harris, E. G. 1991 Am. J. Phys. 59, 421

\bibitem{} T. W. Murphy, K. Nordtvedt \& S. G. Turyshev, 2007 Phys. Rev.
    Lett., 98, 071102   gr-qc/0702028

\bibitem{} I. Ciufolini, arXiv:0704.3338v1 [gr-qc]

\bibitem{} S. Kopeikin, Phys. Rev. Lett. in press, arXiv:gr-qc/0702120v2

\bibitem{} K. Nordtvedt 1999 Class. Quantum Grav. 16, A101-A112

\bibitem{} K. Nordtvedt gr-qc/0301024

\bibitem{} B. Shahid-Saless, 1992 Phys. Rev. D, 46, 5404

\end{thebibliography}
\end{document}